\documentclass[preprint,prd,aps,article,nofootinbib]{revtex4}
\usepackage{graphicx}
\usepackage{diagbox}
\usepackage{float}
\usepackage[colorlinks,
            linkcolor=red,
            anchorcolor=blue,
            citecolor=black
            ]{hyperref}
\usepackage{doi}
\usepackage{color}
\usepackage{slashed}
\begin{document}
\title{ Partial wave effects in the heavy quarkonium radiative electromagnetic decays}
\author{Su-Yan Pei,$^{1,2,}$\footnote{peisuya@163.com}
Wei Li,$^{1,2}$
Ting-Ting Liu,$^{1,2}$
Meng Han,$^{1,2}$
Guo-Li Wang,$^{1,2,}$\footnote{wgl@hbu.edu.cn, corresponding author}
and Tian-Hong Wang$^{3,}$\footnote{thwang@hit.edu.cn,corresponding author}}

\affiliation{$^1$ Department of Physics and Technology, Hebei University, Baoding 071002, China
\nonumber\\
$^{2}$ Key Laboratory of High-precision Computation and Application
of Quantum Field Theory of Hebei Province, Baoding, China\nonumber\\
$^3$ School of Physics, Harbin Institute of Technology, Harbin 150001, China}
\begin{abstract}
In a previous paper \cite{Bc}, it was pointed out that the wave functions of all particles are not pure waves, besides the main partial waves, they all contain {other partial waves}. It is very interesting to know what role these different partial waves play in particle transitions. Therefore,
by using the Bethe-Salpeter equation method, we study the radiative electromagnetic decays $\psi\rightarrow\gamma\chi_{_{cJ}}$ and $\Upsilon\rightarrow\gamma\chi_{_{bJ}}$ ($J=0,1,2$).
We find that for the $S$ and $P$ wave dominated states, like the $\psi(2S)$, $\Upsilon(2S)$, $\chi_{_{cJ}}(1P)$, and $\chi_{_{bJ}}(1P)$ etc., the dominant $S$ and $P$ waves provide main and nonrelativistic contrition to the decays; other partial waves mainly contribute to the relativistic correction. For the states like the $\psi(1D)$, $\Upsilon(2D)$, $\chi_{c2}(1F)$, and $\chi_{b2}(1F)$ etc., they are the $S-P-D$ mixing state dominated by $D$ wave or the $P-D-F$ mixing state dominated by $F$ wave. Large decay widths are found in the transitions $\psi(2D)\to \chi_{c2}(1F)$, $\Upsilon(1D)\to \chi_{bJ}(1P)$, and $\Upsilon(2D)\to \chi_{bJ}(2P)$ etc., which may be helpful to study the missing states $\chi_{c2}(1F)$, $\Upsilon(1D)$, and $\Upsilon(2D)$.
\end{abstract}
\maketitle

\section{Introduction}

In a previous paper \cite{Bc}, we pointed out that the wave functions of all particles are not pure waves. For example, the wave function of a $0^-$ state is a $S$ wave dominant, but it also contains a small amount of $P$ partial wave, which is a relativistic correction to the nonrelativistic $S$ wave. For the $1^-$ case, there are two kinds of states. One is dominated by the $S$ wave and contains small amounts of $P$ and $D$ waves, which are relativistic corrections. The other is dominated by the $D$ wave and contains $S$ and $P$ wave components, where the $P$ wave  can be simply treated as a relativistic correction, but the $S$ wave is somewhat complex, as it is the subtraction of two large terms. Although its total contribution is small in this article, it cannot be simply treated as a relativistic correction.

In order to know the behavior of each component of a wave function in particle transition, we study the radiative electromagnetic decays of heavy quarkonium, namely the transition $1^{--}\to 0^{++},~1^{++},~2^{++}$, which includes the decays of $\psi(nS)\rightarrow\gamma\chi_{_{cJ}}(mP)$, $\Upsilon(nS)\rightarrow\gamma\chi_{_{bJ}}(mP)$ ($n=2,3$, $m=1,2$), $\psi(n'D)\rightarrow\gamma\chi_{_{cJ}}(m'P)$, $\Upsilon(n'D)\rightarrow\gamma\chi_{_{bJ}}(m'P)$ ($n'=1,2$, $m'=1,2$), $\psi(2D)\rightarrow\gamma\chi_{_{c2}}(1F)$, and $\Upsilon(2D)\rightarrow\gamma\chi_{_{b2}}(1F)$ etc., where $J=0,1,2$.
We will first give the contents of different partial waves in the wave functions of the heavy quarkonia involved and then show the details of the contributions of each partial wave in the decay process.

Heavy quarkonia have attracted great interest both in theory and experiment since they were discovered \cite{ret1,ret1-2,bb1,bb2}.
So far, great progress has been made in experiment \cite{ret2,ret4,ret6,ret36,ret32,ret12,ret30,nora} and theory \cite{ret23,ret21,ret22,ret16,ret27,ret3,ret40,ret37,ret5,chen1,chen2}.
However, there are still some radiative decays that have not been detected experimentally, for example, $\psi(4040)\to \chi_{cJ}(2P)\gamma$ and $\psi(4160)\to \chi_{cJ}(2P)\gamma$ ($J=0,1,2$), etc. There are even some particles that have not been discovered such as the $\Upsilon(1D)$, $\Upsilon(2D)$, $\Upsilon(1F)$, and $\psi(1F)$, etc. Therefore, the study of the radiative transitions will help to detect the missing channels and discover the missing particles experimentally.

In this paper, we choose the Bethe-Salpeter (BS) equation method. The BS equation \cite{ret7} is a relativistic dynamic equation dealing with bound states in quantum field theory. The Salpeter equation \cite{ret8} is its instantaneous approximation, which is suitable for heavy quarkonia. Since we solve the complete Salpeter equation without other approximations, we can obtain the meson wave function which contains components of multiple partial waves. This method has been proved to have good results in many aspects \cite{ret9,ret13,ret11,liwei,puzzle,vn,ee}.

This paper is organized as follows. In Sec. II, we will show the method to calculate the transition matrix element of the EM decay. The emphasis is given on the partial waves of the wave functions. In Sec. III, we give the results of the EM decay and discuss the contribution of different partial waves to the decay width.

\section{THE ORETICAL METHOD}
\subsection{Transition matrix element}

The transition matrix for the single-photon radiative decay of a $1^{--}$ quarkonium can be written as
\begin{eqnarray}
&&<P_{_{f}};k,\epsilon_{_0}|S|P,\epsilon>=(2\pi)^{4}ee_{_{q}}\delta^{4}(P-P_{_{f}}-k)\epsilon_{_{0}}^{\mu}M_{_{\mu}},
\end{eqnarray}
where $P$, $P_{f}$, and $k$ are the momenta of the initial meson, the final meson and the final photon, respectively; $ee_{q}$ are the charge of the quark, $e_q=2/3$ and $1/3$ for the charm and bottom quarks, respectively; $\epsilon$ and $\epsilon_{_0}$ are the polarization vectors of the initial quarkonium and the final photon, respectively.


In the BS equation method,
the transition amplitude $M^{\mu}$ can be expressed as the overlapping integral over the initial and final state wave functions,
\begin{eqnarray}
M^{\mu}=&&\int\frac{d^{4}qd^{4}q_{_{1}}}{(2\pi)^{4}}\delta^{4}(p_{_{2}}-p_{_{f2}})Tr\left[\overline{\chi}_{_{P_{_{f}}}}(q_{_{1}})\gamma^{\mu}\chi_{_{P}}(q)S_{_{2}}^{-1}(-p_{_{2}})\right]
\nonumber\\
       &&-\int\frac{d^{4}qd^{4}q_{_{2}}}{(2\pi)^{4}}\delta^{4}(p_{_{1}}-p_{_{f1}})Tr\left[\overline{\chi}_{_{P_{_{f}}}}(q_{_{2}})S_{_{1}}^{-1}(p_{_{1}})\chi_{_{P}}(q)\gamma^{\mu}\right],
\end{eqnarray}
where the subscript $f$ means the final state; $\chi_{_{P}}$ and $\overline{\chi}_{_{P_{_{f}}}}$ are, respectively, the BS wave
functions of the initial and final mesons with $\overline{\chi}=\gamma_0\chi^{\dag}\gamma_0$. $S_{_{1}}(p_{_{1}})$ and $S_{_{2}}(-p_{_{2}})$ are the propagators of quark $1$ with momentum $p_{_1}$ and antiquark $2$ with momentum $p_{_2}$, respectively; $q$, $q_{_1}$, and $q_{_2}$ are the relative momentum between quark and antiquark in the mesons, for example, $q=p_{_1}-\frac{1}{2}P=\frac{1}{2}P-p_{_2}$. The final state relative momentum $q_{_1}$ or $q_{_2}$ is related to the initial relative momentum $q$ by the $\delta$-function for the spectator, so we have the relations $q_{_1}=q+\frac{1}{2}(P_{_f}-P)$ and $q_{_2}=q+\frac{1}{2}(P-P_{_f})$.

Since both the charmonium and bottomonium are double heavy mesons, the instantaneous approximation is a good choice to avoid solving the complicated full BS equation. With instantaneous approximation, the BS equation becomes the Salpeter equation, and the BS wave function becomes the Salpeter wave function. Under the instantaneous approximation condition, the interaction kernel between quark and antiquark satisfies the following relationship \cite{ret8}: $V(P,q)= V(M, q_{\bot})$ (where M and $q$ are the mass of the meson and relative momentum between quark and antiquark, respectively; $q_{_{\bot}}=q-\frac{q\cdot P}{M^2}P$). In our calculation, the Cornell potential, which is the superposition of linear scalar potential $V_{s}(r)=\lambda r$ and vector potential $V_{v}(r)=-\frac{4}{3}\frac{\alpha_{s}(r)}{r}$, is chosen as the interaction kernel \cite{ret9}: $V(r)=\lambda r+V_{0}-\gamma_{0}\bigotimes\gamma^{0}\frac{4}{3}\frac{\alpha_{s}(r)}{r}$, where $\lambda$ is the string constant, $\alpha_{s}(r)$ is the running coupling constant, and $V_{0}$ is a constant in potential model.

After using the instantaneous approximation, the transition amplitude can be written as \cite{ret11},
\begin{eqnarray}\label{trans}
M^{\mu}=&&\int\frac{d^{3}q_{_{\bot}}}{(2\pi)^{3}}Tr\left[\frac{\slashed{P}}{M}\overline{\varphi}_{_{P_{_{f}}}}^{++}(q_{_{1\bot}})\gamma^{\mu}\varphi_{_{P}}^{++}(q_{_{\bot}})\right]
-\int\frac{d^{3}q_{_{\bot}}}{(2\pi)^{3}}Tr\left[\overline{\varphi}_{_{P_{_{f}}}}^{++}(q_{_{2\bot}})\frac{\slashed{P}}{M}\varphi_{_{P}}^{++}(q_{_{\bot}})\gamma^{\mu}\right],
\end{eqnarray}
where  $M$ is the mass of initial meson. In the center of mass frame of the initial state, we have $q_{_{\bot}}=(0,\vec{q})$. Note that in Eq.(\ref{trans}), we only keep the dominant contribution of positive energy wave function $\varphi^{++}$ but ignore other tiny contributions from negative wave functions, etc.

\subsection{The positive wave functions and their partial waves}

The wave function will be given in the center of mass system of the corresponding meson. $q_{_\bot}$ is the relative momentum between quark and antiquark. $P$ and $M$ are the momentum and mass of the meson, respectively. We know that a meson is generally represented by $^{2S+1}L_J$ or $J^{P(C)}$, where $J$, $S$, and $L$ represent the total angular moment, the spin, and the orbital angular momentum; the parity and charge conjugate parity are general calculated as $P=(-1)^{L+1}$ and $C=(-1)^{L+S}$. In previous paper \cite{Bc}, we point out that the representations of $^{2S+1}L_J$, $P=(-1)^{L+1}$, and $C=(-1)^{L+S}$ are generally suitable for a nonrelativistic wave function, not for a relativistic one, since in a relativistic condition, sometimes $S$ and $L$ are no longer good quantum numbers. We also know that the representation of $J^{P(C)}$ is correct in any case, so in our method, the form of the relativistic wave function is given according to the $J^{P(C)}$ of the corresponding meson, that is, each term in the wave function has the same $J^{P(C)}$ with that of the meson. In the paper \cite{Bc}, the details of how to calculate the parity $P$ and charge conjugate parity $C$ of the wave function are given, for example, after performing the transformations $(P_0,\vec{P})\to (P_0,-\vec{P})$ and $(q_0,\vec{q})\to (q_0,-\vec{q})$ in the wave function, then the value of parity $P$ is obtained.
\subsubsection{$1^{--}$ quarkonium}
The positive energy wave function of the initial $1^{--}$ heavy quarkonium can be expressed as \cite{ret18},
\begin{eqnarray}\label{1-}
\varphi_{1^{--}}^{++}(q_{_\bot})=&&(\epsilon\cdot q_{_\bot})\left[A_{1}+\frac{\slashed{P}}{M}A_{2}+\frac{\slashed{q}_{\bot}}{M}A_{3}+\frac{\slashed{P}\slashed{q}_{\bot}}{M^{2}}A_{4}\right]
\nonumber\\
&&+M\slashed{\epsilon}\left[A_{5}+\frac{\slashed{P}}{M}A_{6}+\frac{\slashed{P}\slashed{q}_{\bot}}{M^{2}}A_{7}\right],
\end{eqnarray}
where $\epsilon^{\mu}$ is the polarization vector of the $1^{--}$ quarkonium; $A_i$ ($i=1,2,...7$) is related to the four independent radial wave functions $a_3$, $a_4$, $a_5$, and $a_6$, which are functions of $-q_{_\bot}^2$ and their numerical values are solutions of full Salpeter equation for $1^{--}$ state,
$$f=\frac{1}{2}\left(a_3+\frac{m}{w}a_4\right),~~
A_{1}=\frac{q_{_\bot}^{2}}{Mm}f+\frac{M}{2m}\left(a_{5}-\frac{m}{w}a_{6}\right),~~
A_{5}=\frac{1}{2}\left(a_{5}-\frac{w}{m}a_{6}\right),~~$$
$$A_{2}=-\frac{M}{w}A_5,~~
A_{3}=f-\frac{M^{2}}{2mw}a_{6},~~
A_{4}=\frac{w}{m}f-\frac{M^{2}}{2mw}a_{5},~~
A_{6}=-\frac{m}{w}A_5,~~
A_{7}=A_2,$$
where $m$ and $w=\sqrt{m^2-q_{_\bot}^2}$ are the mass and energy of the quark (antiquark), respectively. We can see that \cite{Bc}, each term in Eq. (\ref{1-}) has the quantum number of $J^{PC}=1^{--}$.

Reference \cite{Bc} pointed out that the wave function of the $1^-$ $B^*_c$ state is not a pure $S$ wave; it includes $P$ and $D$ partial waves \cite{Bc}. The same conclusion is applicable to the $1^{--}$ quarkonium. In the wave function of Eq. (\ref{1-}), the terms including $A_5$ and $A_6$ are $S$ waves, the $A_1$, $A_2$, and $A_7$ terms are $P$ waves, while $A_3$ and $A_4$ terms are $D$ waves mixed with $S$ waves since
$$(\epsilon\cdot q_{_\bot})\slashed{q}_{\bot}=\frac{1}{3}q_{_\bot}^{2}\slashed{\epsilon}+\left[(\epsilon\cdot q_{_\bot})\slashed{q}_{\bot}-\frac{1}{3}q_{_\bot}^{2}\slashed{\epsilon}\right],$$
where $\frac{1}{3}q_{_\bot}^{2}\slashed{\epsilon}$ is $S$ wave, $(\epsilon\cdot q_{_\bot})\slashed{q}_{\bot}-\frac{1}{3}q_{_\bot}^{2}\slashed{\epsilon}$ is $D$ wave.
So in Eq. (\ref{1-}), the complete $S$ wave expression is
\begin{eqnarray}\label{pure1-s}
M\slashed{\epsilon}\left[A_{5}+\frac{\slashed{P}}{M}A_{6}\right]
+\frac{1}{3}q_{_\bot}^{2}\slashed{\epsilon}\left[\frac{1}{M}A_{3}
-\frac{\slashed{P}}{M^{2}}A_{4}\right],
\end{eqnarray}
and $D$ wave is
\begin{eqnarray}\label{pure1-d}
\left[\epsilon\cdot q_{_\bot}\slashed{q}_{\bot}-\frac{1}{3}q_{_\bot}^{2}\slashed{\epsilon}\right]\left[\frac{1}{M}A_{3}
-\frac{\slashed{P}}{M^{2}}A_{4}\right].
\end{eqnarray}

\begin{table}[h]
\begin{center}
\caption{Ratios of the partial waves in the $1^{--}$ wave functions for heavy quarkonia.}
\label{spd}
\resizebox{\linewidth}{!}{
\begin{tabular}{|c|c|c|c|c|c|c|}
\hline
              &               & 1 $S$                & 2 $S$              &  1 $D$              &3 $S$               &2 $D$                \\ \hline
~$\psi$~      &~$S~:~P~:~D$ ~ &1~:~0.126~:~0.0551    &1~:~0.148~:~0.0647  &0.0631~:~0.171~:~1   &1~:~0.170~:~0.0705  &0.0711~:~0.199~:~1   \\ \hline
~$\Upsilon ~$ &$S~:~P~:~D$    &1~:~0.0395~:~0.0205   &1~:~0.0434~:~0.0236 &0.0202~:~0.0443~:~1  &1~:~0.0486~:~0.0259 &0.0244~:~0.0495~:~1  \\ \hline
\end{tabular}}
\end{center}
\end{table}

After solving the full Salpeter equation for the $1^{--}$ state \cite{ret18}, the numerical results indicate that the solutions contain two types of results: one is $S$ wave dominant states, and the other is $D$ wave dominant states. To see the details, following the method of Ref. \cite{Bc}, we calculate the ratios between different partial waves in wave functions for the first five states, and the results are shown in Table \ref{spd} where we can see that all the $1^{--}$ states are $S-P-D$ mixing states. The first, second, and fourth solutions corresponding to the $J/\psi$, $\psi(3686)$, and $\psi(4040)$, are dominated by the $S$ partial waves, so they are marked as $\psi(1S)$, $\psi(2S)$, and $\psi(3S)$ in Table \ref{spd}. The third and fifth results corresponding to $\psi(3770)$ and $\psi(4160)$ indicate that the components of $D$ waves are the dominant ones in their wave functions, so they are marked as $\psi(1D)$ and $\psi(2D)$ in Table \ref{spd}. We obtain similar conclusions for the bottomonia.

In a previous paper \cite{Bc}, we use an approximate method, counting the number of $q_{_{\perp}}$, to distinguish different partial waves, such as the zero-$q_{_{\perp}}$ term is the $S$ wave, the one-$q_{_{\perp}}$ term is the $P$ wave, the two-$q_{_{\perp}}$ term is the $D$ wave, and the three-$q_{_{\perp}}$ term is the $F$ wave. This method is suitable for states dominated by a $S$ or $P$ wave, but not for states dominated by a $D$ or $F$ wave. For example, Eq. (\ref{pure1-s}) indicates that a $S$ wave has two sources, one is from $A_5$ and $A_6$ terms (zero-$q_{_{\perp}}$), the other is from $A_3$ and $A_4$ terms (two-$q_{_{\perp}}$), and the two parts of the $S$ wave are in a dissipative relationship. Table \ref{spd} shows that, for a $S$ wave dominated state, the content of the $D$ wave (from $A_3$ and $A_4$ terms) is small, which means that the $S$ wave content from $A_3$ and $A_4$ terms is also small, and the dominant $S$ wave is mainly from $A_5$ and $A_6$ terms. While for a $D$ wave dominated state, the large $D$ wave content means the $S$ wave content from $A_3$ and $A_4$ terms is also high, however, we find that the $S$ wave content from $A_5$ and $A_6$ terms is also large, resulting in a not large content of overall $S$ wave in the wave function.

In the nonrelativistic limit, for a $S$ dominant state, only the $S$ wave from the $A_5$ and $A_6$ terms has a contribution; both the $P$ and $D$ partial waves, as well as the $S$ wave from $A_3$ and $A_4$ terms, disappear as they are all relativistic corrections. While for a $D$ wave dominant state, such as $\psi(3770)$, the situation is completely different; in a nonrelativistic limit, both the $D$ wave and $S$ wave from $A_3$ and $A_4$ terms, as well as the $S$ wave from $A_5$ and $A_6$ terms survive, and only the $P$ partial wave disappears, which corresponds to the relativistic correction. In Table \ref{spd}, taking the $1S$ states as examples, we can see that the content of the $P$ or $D$ partial wave in the wave function of $J/\psi$ is much larger than those of $\Upsilon(1S)$, which means the relativistic correction of $J/\psi$ is much larger than that of $\Upsilon(1S)$.
\subsubsection{$0^{++}$ quarkonium}
The positive energy wave function of the $0^{++}$ state is expressed as \cite{ret19}
\begin{eqnarray}\label{0+wave}
\varphi_{0^{++}}^{++}(q_{_\bot})=&&\slashed{q}_{\bot}B_{1}+\frac{\slashed{P}\slashed{q}_{\bot}}{M}B_{2}+B_{3},
\end{eqnarray}
with
$$B_{1}=\frac{1}{2}(b_{1}+\frac{m}{w}b_{2}),~~
B_{2}=\frac{w}{m}B_1,~~
B_{3}=\frac{q_{\bot}^{2}}{m}B_1,
$$
where $b_1$ and $b_2$ are functions of $-q_{_\bot}^2$, and they are two independent radial parts of the $0^{++}$ wave function, which will be obtained numerically by solving the Salpeter equation for a $0^{++}$ state \cite{ret19}. In Eq.(\ref{0+wave}), the $B_1$ and $B_2$ terms are $P$ waves, while the $B_3$ term is $S$ wave. We show the partial wave ratios of $P:S$ for $\chi_{c0}$ and $\chi_{b0}$ in Table \ref{ps}; we can see that they are all $P$ wave dominant states. The $P$ wave provides nonrelativistic contribution, while the $S$ partial wave, the relativistic correction.

\begin{table}[h]
\begin{center}
\caption{Ratios of the partial waves in the $0^{++}$ and $1^{++}$ wave functions for heavy quarkonia.}
\label{ps}
{\begin{tabular}{|c|c|c|c|c|}\hline
                &            & 1 $P$           & 2 $P$            &  3 $P$          \\ \hline
~~$\chi_{c0}$~~ &~~$P~:~S$~~ &~~1 ~:~0.127 ~~  & ~~ 1~:~0.142 ~~  &~~ 1~:~0.157 ~~  \\ \hline
$\chi_{b0}$     & $P~:~S$    &~~1 ~:~0.0361 ~~ &~~ 1~:~0.0399 ~~  &~~ 1~:~ 0.0444~~ \\ \hline
$\chi_{c1}$     &$P~:~D$     &~~1 ~:~0.137 ~~  &~~ 1~:~0.150 ~~   &~~ 1~:~0.165 ~~  \\ \hline
$\chi_{b1}$     &$P~:~D$     &~~1 ~:~0.0358 ~~ &~~1 ~:~0.040 ~~   &~~1 ~:~0.0463 ~~ \\ \hline
\end{tabular}}
\end{center}
\end{table}
\subsubsection{$1^{++}$ quarkonium}
The positive energy wave function of $1^{++}$ state is expressed as \cite{ret19}
\begin{eqnarray}\label{1+}
\varphi_{1^{++}}^{++}(q_{_\bot})=i\varepsilon_{\mu\nu\alpha\beta}P^{\nu}q_{_\bot}^{\alpha}\epsilon^{\beta}\gamma^{\mu}\left[D_{1}M+D_{2} \slashed{P}
+D_{3}\slashed{P}\slashed{q}_{\bot}\right]/M^2,
\end{eqnarray}
with
$$D_{1}=\frac{1}{2}(d_{1}+\frac{w}{m}d_{2}),~~
D_{2}=-\frac{m}{w}D_1,~~
D_{3}=-\frac{1}{w}D_1,
$$
where $\varepsilon_{\mu\nu\alpha\beta}$ is the Levi-Civita tensor and $\epsilon^{\beta}$ is the polarization vector of the $1^{++}$ state. The two radial wave functions $d_1$ and $d_2$ are solutions of the corresponding Salpeter equation \cite{ret19}. In Eq. (\ref{1+}), the terms including $D_1$ and $D_2$ are $P$ waves, while the $D_3$ term is a $D$ wave. We obtain the ratios of $P:D$ for $\chi_{c1}$ and $\chi_{b1}$ and show them in Table \ref{ps}. The results indicate that they are all $P$ wave dominant states; that is, $P$ wave provides the nonrelativistic contribution, and the $D$ wave provides the relativistic correction.
\subsubsection{$2^{++}$ quarkonium}
The positive energy wave function of $2^{++}$ state is expressed as \cite{ret20}
\begin{eqnarray}\label{2+}
\varphi_{2^{++}}^{++}(q_{_\bot})=&&\epsilon_{\mu\nu}q_{_\bot}^{\nu}q_{_\bot}^{\mu}\left[F_{1}+\frac{\slashed{P}}{M}F_{2} +\frac{\slashed{q}_{\bot}}{M}F_{3}+\frac{\slashed{P}\slashed{q}_{\bot}}{M^{2}}F_{4}\right]
\nonumber\\
&&+M\epsilon_{\mu\nu}\gamma^{\mu}q_{_\bot}^{\nu}\left[F_{5}+\frac{\slashed{P}}{M}F_{6} +\frac{\slashed{P}\slashed{q}_{\bot}}{M^{2}}F_{7}\right],
\end{eqnarray}
where $\epsilon_{\mu\nu}$ is the polarization tensor of the $2^{++}$ state. $F_i$ ($i=1,2,...,7$) is a function of the four independent radial wave functions $f_3$, $f_4$, $f_5$, and $f_6$, which are solutions of the Salpeter equation for $2^{++}$ state \cite{ret20},
$$
g=\frac{1}{2}\left(f_3+\frac{m}{w}f_4\right),~~
F_{1}=\frac{q_{_\bot}^{2}}{Mm}g+\frac{M}{2m}\left(f_{5}-\frac{m}{w}f_{6}\right),~~
F_{5}=\frac{1}{2}\left(f_{5}-\frac{w}{m}f_{6}\right),~~
$$
$$
F_{2}=-\frac{M}{w}F_5,~~
F_{3}=g-\frac{M^{2}}{2mw}f_{6},~~
F_{4}=\frac{w}{m}g-\frac{M^{2}}{2mw}f_{5},~~
F_{6}=-\frac{m}{w}F_5,~~
F_{7}=F_2.
$$

In the wave function of Eq. (\ref{2+}), $F_5$ and $F_6$ terms are $P$ waves, $F_1$, $F_2$, and $F_7$ terms are $D$ waves, while $F_3$ and $F_4$ terms include both $P$ and $F$ waves, since
$$
\epsilon_{\mu\nu}q_{_\bot}^{\nu}q_{_\bot}^{\mu}\slashed{q}_{\bot}=
\frac{2}{5}q_{_\bot}^{2}\epsilon_{\mu\nu}q_{_\bot}^{\nu}\gamma^{\mu}
+\left[\epsilon_{\mu\nu}q_{_\bot}^{\nu}q_{_\bot}^{\mu}\slashed{q}_{\bot}-\frac{2}{5}q_{_\bot}^{2}\epsilon_{\mu\nu}q_{_\bot}^{\nu}\gamma^{\mu}\right],
$$
where $\frac{2}{5}q_{_\bot}^{2}\epsilon_{\mu\nu}q_{_\bot}^{\nu}\gamma^{\mu}$ is a $P$ wave, and $\epsilon_{\mu\nu}q_{_\bot}^{\nu}q_{_\bot}^{\mu}\slashed{q}_{\bot}-\frac{2}{5}q_{_\bot}^{2}\epsilon_{\mu\nu}q_{_\bot}^{\nu}\gamma^{\mu}$ is a $F$ wave. So in Eq. (\ref{2+}), the $F$ wave is
\begin{eqnarray}\label{pure2+f}
\left[\epsilon_{\mu\nu}q_{_\bot}^{\nu}q_{_\bot}^{\mu}\slashed{q}_{\bot}
-\frac{2}{5}q_{_\bot}^{2}\epsilon_{\mu\nu}q_{_\bot}^{\nu}\gamma^{\mu}\right]\left[\frac{1}{M}F_{3}
-\frac{\slashed{P}}{M^{2}}F_{4}\right],
\end{eqnarray}
and the overall $P$ wave is
\begin{eqnarray}\label{pure2+p}
M\epsilon_{\mu\nu}\gamma^{\mu}q_{_\bot}^{\nu}\left[F_{5}
+\frac{\slashed{P}}{M}F_{6}\right]
+\frac{2}{5}q_{_\bot}^{2}\epsilon_{\mu\nu}q_{_\bot}^{\nu}\gamma^{\mu}\left[\frac{1}{M}F_{3}
-\frac{\slashed{P}}{M^{2}}F_{4}\right].
\end{eqnarray}

The ratios $P:D:F$ of partial waves for some $2^{++}$ heavy quarkonia are shown in Table \ref{pdf}. Where we can see, all the $2^{++}$ quarkonia are $P-D-F$ mixing states. The first, second, and fourth solutions correspond with the $P$ wave dominant states, since in the nonrelativistic limit, only the $P$ wave from $F_5$ and $F_6$ terms has a contribution; the $D$ wave, $F$ wave, and $P$ wave from $F_3$ and $F_4$ terms are all disappear, so these $P$ wave dominant states are marked as $1P$, $2P$, and $3P$ states in Table \ref{pdf}. The third and fifth solutions corresponding to the $F$ wave dominant states are marked as the $1F$ and $2F$ states. In a nonrelativistic limit, both the $F$ wave and $P$ wave from $F_3$ and $F_4$ terms, as well as the $P$ wave from $F_5$ and $F_6$ terms have contributions; only the $D$ wave disappears, which corresponds to the relativistic correction.

\begin{table}[h]
\begin{center}
\caption{Ratios of the partial waves in the $2^{++}$ wave functions for heavy quarkonia.}
\label{pdf}
\resizebox{\linewidth}{!}{
{\begin{tabular}{|c|c|c|c|c|c|c|}
\hline
               &                 & 1 $P$             & 2 $P$               &  1 $F$               & 3 $P$               & 2 $F$             \\ \hline
~$\chi_{c2}$~  &~$P~:~D~:~F$  ~  &1~:~0.146~:~0.0542 &1~:~0.166~:~0.0670   &0.0655~:~0.194~:~1    &1~:~0.189~:~0.0701   &0.0720~:~0.206~:~1 \\ \hline
~$\chi_{b2}$  ~&$P~:~D~:~F$      &1~:~0.0411~:~0.0199&1~:~0.0478~:~0.0249  &0.0216~:~0.0482~:~1   &1~:~0.0533~:~0.0278  &0.0260~:~0.0541~:~1\\ \hline
\end{tabular}}}
\end{center}
\end{table}

\subsection{Form factors}
With the positive energy wave functions, the transition amplitude in Eq. (\ref{trans}) are calculated straightly, and they are expressed as functions of the form factors. For the transitions of $1^{--} \to 0^{++},1^{++},2^{++}$, the transition amplitudes can be written as
\begin{eqnarray}
M^{\mu}_{_{1^{--}\rightarrow0^{++}}}=&&P^{\mu}(\epsilon\cdot P_{_{f}})t_{_{1}}
+\epsilon^{\mu}t_{_{2}},
\nonumber\\
M^{\mu}_{_{1^{--}\rightarrow1^{++}}}=&&P^{\mu}\epsilon^{\epsilon \epsilon_{_{f}} P P_{_{f}}}x_{_{1}}
+(\epsilon\cdot P_{_{f}})\epsilon^{\mu \epsilon_{_{f}} P P_{_{f}}}x_{_{2}}+\epsilon^{{\mu}\epsilon \epsilon_{_{f}} P_{_{f}}}x_{_{3}},
\nonumber\\
M^{\mu}_{_{1^{--}\rightarrow2^{++}}}=&&P^{\mu}(\epsilon\cdot P_{_{f}})\epsilon_{_{f}}^{P P}y_{_{1}}+\epsilon^{\mu}\epsilon_{_{f}}^{P P}y_{_{2}}+P^{\mu}\epsilon_{_{f}}^{\epsilon P}y_{_{3}}
+(\epsilon\cdot P_{_{f}})\epsilon_{_{f}}^{\mu P}y_{_{4}}+\epsilon_{_{f}}^{\mu \epsilon}y_{_{5}},
\end{eqnarray}
where $t_{_{1}}$, $t_{_{2}}$, $x_{_{i}}$ $(i=1,2,3)$ and $y_{_{j}}$ $(j=1,2,...5)$ are the form factors; $P_{_{f}}$ is the momentum of the final state; and $\epsilon_{_{f}}^{\mu}$ or $\epsilon_{_{f}}^{\mu \nu}$ is the polarization vector or tensor of the final quarkonium. In the upper formula, we have used the following shorthand notations: $\epsilon_{_{f}}^{P P}\equiv \epsilon_{_{f}}^{\mu\nu}P_{\mu}P_{\nu}$ and $\epsilon^{\epsilon \epsilon_{_{f}} P P_{_{f}}}=\epsilon^{\rho\sigma\alpha\beta}\epsilon_{_\rho} \epsilon_{_{f\sigma}} P_{_\alpha} P_{_{f\beta}}$, etc.

In the transition amplitude, because of the relation $k\cdot\epsilon_{_0}=0$, where $k=P-P_{_f}$ and $\epsilon_{_0}$ are the momentum and polarization vector of the photon, some of the expressions do not exist at the same time. For example, in the transition $1^{--}\rightarrow 0^{++}$, there is the term of $P^{\mu}
(\epsilon\cdot P_{_{f}})$, but no $P_{_{f}}^{\mu}(\epsilon\cdot P_{_{f}})$ term. For the same reason, there are no $P_{_{f}}^{\mu}$ terms in the transitions of $1^{--}\rightarrow 1^{++}$ and $1^{--}\rightarrow 2^{++}$. Since the transition amplitude must meet the gauge invariance, we also have the following relations:
\begin{eqnarray}
t_{_{2}}=&&M(M-E_{_{f}})t_{_{1}},
\nonumber\\
x_{_{3}}=&&-M(M-E_{_{f}})x_{_{1}},
\nonumber\\
y_{_{4}}=&&y_{_{2}}-M(M-E_{_{f}})y_{_{1}},
\nonumber\\
y_{_{5}}=&&-M(M- E_{_{f}})y_{_{3}}.
\end{eqnarray}

\section{Results and discussions}
\subsection{Charmonium's radiative decays}
The masses of most low excited charmonia have been detected by experiment \cite{PDG},
\begin{eqnarray}
M_{\psi(2S)}=&&3686 ~\rm{MeV},~ M_{\psi(3770)}=3773 ~\rm{MeV},~ M_{\psi(4040)}=4039 ~\rm{MeV},
\nonumber\\
M_{\psi(4160)}=&&4191 ~\rm{MeV},~ M_{\chi_{_{c0}}(1P)}=3414 ~\rm{MeV},~ M_{\chi_{_{c1}}(1P)}=3510 ~\rm{MeV},
\nonumber\\
M_{\chi_{_{c2}}(1P)}=&&3556 ~\rm{MeV},~ M_{\chi_{_{c0}}(3860)}=3860 ~\rm{MeV},~M_{\chi_{_{c1}}(3872)}=3872 ~\rm{MeV},
\nonumber\\
M_{\chi_{_{c2}}(3930)}=&&3922 ~\rm{MeV}.
\end{eqnarray}

To compare, we also show our model predictions \cite{ret14},
\begin{eqnarray}
M_{\psi(2S)}=&&3688~\rm{MeV},~ M_{\psi(3770)}=3779 ~\rm{MeV},~ M_{\psi(4040)}=4057 ~\rm{MeV},
\nonumber\\
M_{\psi(4160)}=&&4111 ~\rm{MeV},~ M_{\chi_{_{c0}}(1P)}=3415 ~\rm{MeV},~ M_{\chi_{_{c1}}(1P)}=3510 ~\rm{MeV},
\nonumber\\
M_{\chi_{_{c2}}(1P)}=&&3556 ~\rm{MeV},~ M_{\chi_{_{c0}}(3860)}=3837 ~\rm{MeV}, ~M_{\chi_{_{c1}}(3872)}=3929 ~\rm{MeV},
\nonumber\\
M_{\chi_{_{c2}}(3930)}=&&3972 ~\rm{MeV}.
\end{eqnarray}
Most of our predictions on mass spectra are in good agreement with experimental data, with a few values that differ by tens of MeV from the experimental data, such as the masses of $\psi(4160)$, $\chi_{_{c1}}(3872)$, and $\chi_{_{c2}}(3930)$. In our calculation, when solving the Salpeter equation for $1^{--}$ state, we choose the experimental masses by adjusting the free parameter $V_0$ in the potential to fit experimental data for each state and obtain the numerical value of corresponding wave function.
However, there are still some particles that have not been detected by experiments, such as $\chi_{_{c2}}(1F)$. For such particles, our theoretical predictions are used; for example, the mass of $\chi_{_{c2}}(1F)$ is predicted as $4037$ MeV \cite{ret14}.

\begin{table}[h]
\begin{center}
\caption{The decay widths (keV) of $\Psi\to\chi_{_{cJ}}\gamma$. Where $\chi_{_{c0}}(3860)$, $\chi_{_{c1}}(3872)$, and $\chi_{_{c2}}(3930)$ are treated as the $2P$ states, $\chi_{_{c2}}(1F)$ is the $1F$ dominant state with sizable $P$ and $D$ partial waves; our theoretical prediction about its mass is 4037 MeV.}\label{charm}
\begin{tabular}{c c c c c c c c c c}
\hline
\textbf{Process}&\textbf{Ours}&\textbf{\cite{ret16}}&\textbf{\cite{ret41}}&\textbf{\cite{ret17}}
&\textbf{\cite{ret42}}&\textbf{\cite{ret34}}&\textbf{\cite{ret31}a}&\textbf{\cite{ret31}b}&
\textbf{PDG}\cite{PDG}  \\
\hline
$\psi(2S)\rightarrow\gamma\chi_{_{c0}}(1P)$      &39.9 &50  &26.3 &47.0&25.2  &26   &22  &22  &$28.8\pm 1.4$    \\
$\psi(2S)\rightarrow\gamma\chi_{_{c1}}(1P)$      &35.6 &45  &22.9 &42.8&29.1  &29   &42  &45  &$28.7\pm 1.5$    \\
$\psi(2S)\rightarrow\gamma\chi_{_{c2}}(1P)$      &24.5 &29  &18.2 &30.1&25.2  &24   &38  &46  &$28.0\pm 1.4$    \\
\hline
$\psi(3770)\rightarrow\gamma\chi_{_{c0}}(1P)$    &290  &    &355  &299 &243.9 &213  &272 &261 &$188\pm23$       \\
$\psi(3770)\rightarrow\gamma\chi_{_{c1}}(1P)$    &90.8 &    &135  &99.0&104.9 &77   &138 &135 &$67.7\pm8.7$     \\
$\psi(3770)\rightarrow\gamma\chi_{_{c2}}(1P)$    &3.50 &    &6.9  &3.88&1.9   &3.3  &7.1 &8.1 &$<17.4$          \\
\hline
$\psi(4040)\rightarrow\gamma\chi_{_{c0}}(1P)$    &0.29 &    &     &    &2.1   &12.7 &5.9 &6.7 &                 \\
$\psi(4040)\rightarrow\gamma\chi_{_{c1}}(1P)$    &1.42 &    &     &    &0.3   &0.85 &4.0 &6.7 &$<272$           \\

$\psi(4040)\rightarrow\gamma\chi_{_{c2}}(1P)$    &2.78 &    &     &    &2.4   &0.63 &0.25&2.5 &$<400$           \\
$\psi(4040)\rightarrow\gamma\chi_{_{c0}}(3860)$  &31.0 &    &     &    &30.1  &22   &19  &27  &                 \\
$\psi(4040)\rightarrow\gamma\chi_{_{c1}}(3872)$  &74.4 &    &     &    &45.0  &43   &55  &67  &                 \\
$\psi(4040)\rightarrow\gamma\chi_{_{c2}}(3930)$  &43.2 &    &     &    &36.0  &48   &67  &82  &                 \\
\hline
$\psi(4160)\rightarrow\gamma\chi_{_{c0}}(1P)$    &8.42 &    &     &    &23.3  &35   &150 &189 &                 \\
$\psi(4160)\rightarrow\gamma\chi_{_{c1}}(1P)$    &6.66 &    &     &    &0.02  &3.4  &37  &63  &                 \\
$\psi(4160)\rightarrow\gamma\chi_{_{c2}}(1P)$    &1.25 &    &     &    &0.23  &0.027&17  &20  &                 \\
$\psi(4160)\rightarrow\gamma\chi_{_{c0}}(3860)$  &462  &    &     &    &      &191  &332 &360 &                 \\
$\psi(4160)\rightarrow\gamma\chi_{_{c1}}(3872)$  &281  &    &     &    &      &114  &309 &347 &                 \\
$\psi(4160)\rightarrow\gamma\chi_{_{c2}}(3930)$  &14.3 &    &     &    &      &6.3  &24  &29  &                 \\
$\psi(4160)\rightarrow\gamma\chi_{_{c2}}(1F)$    &73.9 &    &     &    &      &17   &    &    &                 \\
\hline
\end{tabular}
\end{center}
\end{table}

The theoretical results as well as data from  Particle Data Group \cite{PDG} about the charmonia radiative decays $1^{--}\to 0^{++},1^{++},2^{++}$ are shown in Table \ref{charm}.
$\psi(4040)$ is the state $\psi(3S)$; $\psi(3770)$ and $\psi(4160)$ are the $\psi(1D)$ and $\psi(2D)$; $\chi_{c0}(3860)$, $\chi_{c1}(3872)$, and $\chi_{c2}(3930)$ are the $\chi_{c0}(2P)$, $\chi_{c1}(2P)$, and $\chi_{c2}(2P)$, respectively. Although the masses may be different between models, especially the $\chi_{cJ}(2P)$ states, usually the radiative decay are not very sensitive to the masses except for some special channels, for example, $\psi(4040)\to\chi_{_{cJ}}(1P)\gamma$ and $\psi(4160)\to\chi_{_{cJ}}(1P)\gamma$, which we will discuss later.

At present, only the decays of $\psi(2S)$ and $\psi(3770)$ have experimental results. Our results of them are comparable to experimental data and other theoretical values. For channels $\psi(4040)\to\chi_{_{cJ}}(2P)\gamma$ and $\psi(4160)\to\chi_{_{cJ}}(2P)\gamma$, the values between different theoretical models are also comparable. Compared with the case of bottomonium, where the predictions of different models are in good agreement, the results of charmonium differ slightly between different models. We believe that this is mainly due to the large relativistic correction in the charmonium system.

While for other processes, such as $\psi(4040)\to\chi_{_{cJ}}(1P)\gamma$ and $\psi(4160)\to\chi_{_{cJ}}(1P)\gamma$, there are huge differences between the results predicted by different models.
We find that the huge difference comes from the uncertainty of theory. Take the decay $\psi(4040)\to\chi_{_{cJ}}(1P)\gamma$ as an example to illustrate this. The radial wave function of initial state has two nodes. The contributions of wave functions on both sides of the node to the amplitude are offset. The decay width of this channel is much smaller than that of other processes, which indicates that the cancellation is very strong, leading to the strong dependence of the results on model parameters. Therefore, the theoretical error of this process is huge, and this is the reason that the predictions of different models differ greatly.

\subsection{Contributions of different partial waves in Charmonia decays}
\subsubsection{$\psi(2S) \to \chi_{c0}(1P)\gamma$}
In Sec.II, we show the $\psi(2S)$ is a $2S$ dominant state with small admixtures of $P$ and $D$ partial waves. The ratio of its partial waves is $S : P : D = 1 : 0.148 : 0.0647$. For the $\chi_{c0}(1P)$, its wave function is $1P$ dominant but mixing with a small amount of $S$ wave since we have $P':S'=1:0.127$ (here and later, the superscript ``prime'' is used to denote the partial wave in the final state).
In Table \ref{charm0+}, we show the detailed contributions of the different partial waves to the decay width of $\psi(2S) \to \chi_{c0}(1P)\gamma$. Where the ``whole'' means the result is obtained using the complete wave function, the ``$S$ wave'' in a column or row means the corresponding result is obtained only using the $S$ partial wave and ignoring others, etc.

\begin{table}[H]
\begin{center}
\caption{Contributions of different partial waves to the decay width (keV) of $\psi(2S) \to \chi_{c0}(1P)\gamma$.}\label{charm0+}
{\begin{tabular}{|c|c|c|c|} \hline \diagbox {$1^{--}$}{$0^{++}$}
                        & Whole~($P'+S'$)               & $P'$ wave                      & $S'$ wave                      \\ \hline
  ~~~~~Whole~($S+P+D$)~~~~~ & 39.9                      & 33.6                           & 0.269                          \\ \hline
  $S$ wave              & 34.6                          & 34.6                           & 0                              \\ \hline
  $P$ wave              & 0.215                         & $4.2\times 10^{-3}$            & 0.279                          \\ \hline
  $D$ wave              &~~~~~$8.5\times 10^{-4}$~~~~~  & ~~~~~$3.9\times 10^{-4}$~~~~~  & ~~~~~$9.0\times 10^{-5}$~~~~~  \\ \hline
\end{tabular}}
\end{center}
\end{table}

From Table \ref{charm0+}, we can see that the dominant $S$ partial wave in $\psi(2S)$ state and $P'$ wave in $\chi_{c0}(1P)$ provide the overwhelming contribution. The $P$ and $D$ partial waves in $\psi(2S)$ and $S'$ wave in $\chi_{c0}(1P)$ give a small contribution.

To see the relativistic effect, we calculate the ratio, $$\frac{\Gamma_{rel}-\Gamma_{non-rel}}{\Gamma_{rel}}=13.8\%,$$ where $\Gamma_{rel}$ is the decay width calculated by the whole wave functions Eq.(\ref{1-}) and Eq.(\ref{0+wave}) and $\Gamma_{non-rel}$ is obtained only using the nonrelativistic wave functions. $\psi(2S)$, as a $S$ wave dominant $1^{--}$ state, is a nonrelativistic wave function that only contains $S$ wave from $A_5$ and $A_6$ terms. For $0^{++}$ $\chi_{c0}(1P)$, its nonrelativistic wave function is a $P'$ wave.
In our calculation, the whole wave function is normalized to 1, and the nonrelativistic one is not normalized separately.
The relativistic effect is not as large as we expected, the reason may be due to that the main contribution of the relativistic correction does not come from the interaction $S\times S'$ between the dominant $S$ wave in $\psi(2S)$ with the small $S'$ wave in $\chi_{c0}(1P)$, see Table \ref{charm0+}, or those from $P\times P'$, but from the interaction between the two small terms $P\times S'$. We also note that there is no interaction (zero in Table \ref{charm0+}) between the $S$ partial wave in $\psi(2S)$ with the $S'$ wave in $\chi_{c0}(1P)$, since $S\times S'=0$.

\subsubsection{$\psi(2S) \to \chi_{c1}(1P)\gamma$}

\begin{table}[H]
\begin{center}
\caption{Contributions of different partial waves to the decay width (keV) of $\psi(2S) \to \chi_{c1}(1P)\gamma$.}\label{charm1+}
{\begin{tabular}{|c|c|c|c|}
\hline
\diagbox{$1^{--}$}{$1^{++}$} & Whole ($P'+D'$)& $P'$ wave                     & $D'$ wave                          \\ \hline
  ~~~~~Whole ($S+P+D$)~~~~~  & 35.6           & 27.8                          & 0.484                              \\ \hline
  $S$ wave                   & 30.8           & 26.9                          & 0.131                              \\ \hline
  $P$ wave                   & 0.137          & $4.8\times 10^{-4}$           & 0.122                              \\ \hline
  $D$ wave  &~~~~~$2.7\times 10^{-3}$~~~~~    & ~~~~~$4.4\times 10^{-3}$~~~~~ & ~~~~~$2.1\times 10^{-4}$~~~~~      \\ \hline
\end{tabular}}
\end{center}
\end{table}
The ratio of the partial waves in $\chi_{_{c1}}(1P)$ is $P':D'=1:0.137$, where $P'$ wave provides the nonrelativistic contribution, while the $D'$ wave provides relativistic correction. In Table \ref{charm1+}, we show the details of the decay $\Psi(2S)\to \chi_{_{c1}}\gamma$ where the largest contribution comes from the interaction $S\times P'$ between the dominant partial waves, the $S$ wave in $\Psi(2S)$ and the $P'$ wave in $\chi_{_{c1}}(1P)$. The $P$ wave and $D'$ wave have a small contribution; the $D$ partial wave has a tiny contribution and can be safely ignored.

The relativistic effect is calculated as
$$\frac{\Gamma_{rel}-\Gamma_{non-rel}}{\Gamma_{rel}}={24.5\%},$$
which is about $1.4$ times larger than the those of $\chi_{_{c0}}(1P)$ case.
In this decay, the relativistic corrections come from the interaction $S\times D'$ between the dominant $S$ wave in $\psi(2S)$ and the small $D'$ wave in $\chi_{c1}(1P)$ and also from the interaction $P\times D'$ between the small $P$ wave in $\psi(2S)$ and the $D'$ wave in $\chi_{c1}(1P)$. Their contributions are comparable, $S\times D'\sim P\times D'$; this indicates that the interaction between the later ($P\times D'$) is much stronger than those of the former ($S\times D'$) since the component of $P$ partial wave is 1 order smaller than those of the $S$ wave in $\psi(2S)$. Similar to the case of $\Psi(2S)\to \chi_{_{c0}}\gamma$, the interaction between $P$ waves, $P\times P'$, is very small in the $\Psi(2S)\to \chi_{_{c1}}\gamma$.

\subsubsection{$\psi(2S) \to \chi_{c2}(1P)\gamma$}
The wave function of $\chi_{c2}$ is more complicated than those of $\chi_{c0}(1P)$ and $\chi_{c1}(1P)$. Besides the dominant $P'$ partial wave, it also contains small amounts of $D'$ and $F'$ partial waves, and their ratios are $P':D':F'=1 : 0.146: 0.0642$. In Table \ref{charm2+}, we show the contributions of different partial waves to the decay $\psi(2S) \to \chi_{c2}(1P)\gamma$. We can see that the dominant $S$ and $P'$ partial waves from the initial $\psi(2S)$ and final $\chi_{c2}(1P)$ give main contributions, the $P$ partial wave in $\psi(2S)$ and $D'$ wave in $\chi_{c2}(1P)$ give small contributions, while the contributions from the $D$ wave in $\psi(2S)$ and $F'$ wave in $\chi_{c2}(1P)$ are tiny, which can be ignored safely.

In the nonrelativistic limit, only the $S$ wave from $A_5$ and $A_6$ terms in Eq.(\ref{1-}) and the $P'$ wave from $F_5$ and $F_6$ terms in Eq.(\ref{2+}) have a contribution. The relativistic corrections mainly come from the interactions of $S\times D'$, $P\times D'$, and $P\times P'$. And the relativistic effect in this decay is
$$\frac{\Gamma_{rel}-\Gamma_{non-rel}}{\Gamma_{rel}}=16.4\%.$$

\begin{table}[H]
\begin{center}
\caption{Contributions of different partial waves to the decay width (keV) of $\psi(2S) \to \chi_{c2}(1P)\gamma$.}\label{charm2+}
{\begin{tabular}{|c|c|c|c|c|}
\hline
\diagbox{$1^{--}$}{$2^{++}$} & ~~Whole ($P'+D'+F'$)~~  & $P'$ wave                   & $D'$ wave           & $F'$ wave                   \\ \hline
  ~~~Whole ($S+P+D$)~~~      & 24.5                    & 21.8                        & 0.0496              & $1.2\times 10^{-4}$         \\ \hline
  $S$ wave                   & 21.4                    & 20.1                        & 0.0133              & $1.8\times 10^{-7}$         \\ \hline
  $P$ wave                   & 0.0961                  & 0.0409                    & $6.3\times 10^{-3}$   & $1.3\times 10^{-5}$         \\ \hline
  $D$ wave            &~~~$8.9\times 10^{-5}$~~~ & ~~~$1.4\times 10^{-4}$~~~    & ~~~$1.1\times 10^{-3}$~~~ &~~~ $2.0\times 10^{-4}$~~~  \\ \hline
\end{tabular}}
\end{center}
\end{table}

\subsubsection{$\psi(3770) \to \chi_{c0}(1P)\gamma$}
$\psi(3770)$ is also a $S-P-D$ mixing state, but unlike the $S$ wave dominant $\psi(nS)$ state, it is a $1D$ dominant state. In its wave function, the ratio of different partial waves is calculated as $S : P : D ={0.0631 : 0.171 : 1}$.
In Table \ref{37700+}, we show the contributions of different partial waves to the decay $\psi(3770) \to \chi_{c0}(1P)\gamma$ where the $D$ partial wave in $\psi(3770)$ and $P'$ wave in $\chi_{c0}(1P)$ provide the dominant contribution.

\begin{table}[H]
\begin{center}
\caption{Contributions of different partial waves to the decay width (keV) of $\psi(3770) \to \chi_{c0}(1P)\gamma$.}\label{37700+}
{\begin{tabular}{|c|c|c|c|}
\hline
\diagbox{$1^{--}$}{$0^{++}$}& ~~~Whole ($P'+S'$)~~~       & $P'$ wave               & ~~~~~~~$S'$ wave~~~~~~~                  \\ \hline
  ~~~Whole ($S+P+D$)~~~     & 290                         & 257                     & 0.925                                    \\ \hline
 $D$ wave                   & 255                         & 250              & ~~~~~0.0210~~~~~                                \\ \hline
 $S$ wave      & ~~~~~$4.0\times 10^{-3}$~~~~~& ~~~~~$4.0\times 10^{-3}$~~~~~       & 0                                        \\ \hline
 $P$ wave                   & 0.944                       & 0.0235                  & 0.670                                    \\ \hline
\end{tabular}}
\end{center}
\end{table}

The calculation of relativistic effect is relatively complex. As mentioned earlier, in the wave function of $\psi(3770)$, the $D$ wave is nonrelativistic dominant part, which comes from $A_3$ and $A_4$ terms. Then the contribution of $S$ wave from the same $A_3$ and $A_4$ terms is also large, which can be considered as a nonrelativistic contribution. In addition, from Table \ref{spd}, it can be seen that the proportion of total $S$ wave is small, which means that the individual contribution of $S$ wave from $A_5$ and $A_6$ is comparable with the one from $A_3$ and $A_4$ and can be considered as a nonrelativistic contribution. In this case, the relativistic correction only comes from the $P$ wave, so the relativistic effect is
$$\frac{\Gamma_{rel}-\Gamma_{non-rel}}{\Gamma_{rel}}=13.1\%.$$
If only $D$ wave is nonrelativistic, and both $S$ and $P$ waves are relativistic corrections, then the relativistic effect is $13.8\%$.

\subsubsection{$\psi(3770) \to \chi_{c1}(1P)\gamma$}
Table \ref{37701+} shows the detail contributions of different partial waves to the decay $\psi(3770) \to \chi_{c1}(1P)\gamma$. The $D$ partial wave in $\psi(3770)$ and the $P'$ wave in $\chi_{c1}(1P)$ provide the dominant contribution. Similar to the case of $\psi(3770) \to \chi_{c0}(1P)\gamma$, if the relativistic correction of $\psi(3770)$ only comes from the $P$ wave, the relativistic effect is obtained as
$$\frac{\Gamma_{rel}-\Gamma_{non-rel}}{\Gamma_{rel}}=22.6\%.$$
If both $S$ and  $P$ waves are relativistic corrections, then the relativistic effect is $21.3\%$.

\begin{table}[H]
\begin{center}
\caption{Contributions of different partial waves to the decay width (keV) of $\psi(3770) \to \chi_{c1}(1P)\gamma$.}\label{37701+}
{\begin{tabular}{|c|c|c|c|}
\hline
\diagbox{$1^{--}$}{$1^{++}$}         & ~Whole ($P'+D'$)~              & $P'$ wave             & $D'$ wave                              \\ \hline
  ~~~~~Whole ($S+P+D$)~~~~~          & 90.8                           & 91.8                  & $2.5\times 10^{-3}$                    \\ \hline
  $D$ wave                           & 68.9                           & 71.5                       & 0.026                             \\ \hline
  $S$ wave             & ~~~~~$1.7\times 10^{-5}$~~~~~  & ~~~~~$5.2\times 10^{-3}$~~~~~  & ~~~~~$5.7\times 10^{-3}$~~~~~               \\ \hline
  $P$ wave                           & 1.53                           & 1.45                  & $1.3\times 10^{-3}$                    \\ \hline

\end{tabular}}
\end{center}
\end{table}
\subsubsection{$\psi(3770) \to \chi_{c2}(1P)\gamma$}

Table \ref{37702+} indicates that in the decay of $\psi(3770) \to \chi_{c2}(1P)\gamma$, the main contribution comes from the $D$ wave in $\psi(3770)$ and $P'$ wave in $\chi_{c2}(1P)$. When considering the relativistic effect, the $P'$ wave in $\chi_{c2}(1P)$ provides the nonrelativistic contribution and $D'$ and $F'$ waves gives the relativistic corrections. If the relativistic correction of $\psi(3770)$ only comes from the $P$ wave, the relativistic effect is
$$\frac{\Gamma_{rel}-\Gamma_{non-rel}}{\Gamma_{rel}}=5.98\%.$$
If both $S$ and  $P$ waves are relativistic corrections, then the relativistic effect is $13.5\%$.

\begin{table}[H]
\begin{center}
\caption{Contributions of different partial waves to the decay width (keV) of $\psi(3770) \to \chi_{c2}(1P)\gamma$.}\label{37702+}
{\begin{tabular}{|c|c|c|c|c|}
\hline
\diagbox{$1^{--}$}{$2^{++}$} & Whole        & $P'$ wave                     &$D'$ wave                       & $F'$ wave                           \\ \hline
  ~~~~~Whole~~~~~            & 3.50         & 4.15                          &  0.058                         & $2.1\times 10^{-3}$                 \\ \hline
  ~~~~~$D$ wave~~~~~         & 2.56         & 2.89                          & 0.011                          & $1.4\times 10^{-3}$                 \\ \hline
  $S$ wave & ~~~~~$3.5\times 10^{-3}$~~~~~  & ~~~~~$5.6\times 10^{-3}$~~~~~ & ~~~~~$1.2\times 10^{-5}$~~~~~  & ~~~~~$7.0\times 10^{-7}$~~~~~       \\ \hline
  $P$ wave                   &  0.049       & 0.071                         & 0.020                          & $5.7\times 10^{-5}$                 \\ \hline
\end{tabular}}
\end{center}
\end{table}

\subsubsection{$\psi(4160)\to \chi_{c2}(1F)\gamma$}
In the literature, $\psi(4160)$ is primarily the state $\psi(2D)$ with admixtures of $\psi(3S)$; that is, it is a $3S-2D$ mixing state.
In our method, $\psi(4160)$ is a $2D$ dominant, $S-P-D$ mixing state. Its main radiative decays are the channels of $\psi(4160)\to \chi_{cJ}(2P)\gamma$ ($J=0,1,2$). We will not show the details of these decays since they are similar to those of $\psi(3770)\to \chi_{cJ}(1P)\gamma$.

As the $2D$ dominant state, its mass is heavier than the $1F$ dominant state $\chi_{c2}(1F)$ in our theoretical prediction, so the decay process $\psi(4160)\to \chi_{c2}(1F)\gamma$ exists. This channel is another typical process not encountered before in this article, since the final meson $\chi_{c2}(1F)$ is also a typical mixed state. In the literature, $\chi_{c2}(1F)$ is the $2P-1F$ mixing state, $1F$ dominant but mixed with sizable $2P$ component. But in our method, it is a $1F$ dominant, $P-D-F$ mixing state. $\chi_{c2}(1F)$ is not available in experiment, so for its mass, we use our theoretical prediction, 4038 MeV \cite{ret14}.

Some details of the contributions of different partial waves to the decay channel $\psi(4160)\to \chi_{c2}(1F)\gamma$ are listed in Table \ref{4160}. We note that the main components of the initial and final states, namely, the $D$ wave in $\psi(4160)$ and the $F'$ wave in $\chi_{c2}(1F)$, give the maximum contribution. If the relativistic corrections only come from the $P$ wave in $\psi(4160)$, and $D'$ wave in $\chi_{c2}(1F)$, the relativistic effect is
$$\frac{\Gamma_{rel}-\Gamma_{non-rel}}{\Gamma_{rel}}=21.5\%.$$
If the $S$ and $P$ waves in $\psi(4160)$ and $P'$ and $D'$ waves in $\chi_{c2}(1F)$ are all relativistic corrections, then the relativistic effect is $22.7\%$.

\begin{table}[H]
\begin{center}
\caption{Contributions of different partial waves to the decay width (keV) of $\psi(4160) \to \chi_{c2}(1F)\gamma$, where $\psi(4160)$ and $\chi_{c2}(1F)$ are $2D$ and $1F$ dominant states, respectively.}\label{4160}
\begin{tabular}{|c|c|c|c|c|}
\hline
{\diagbox{$1^{--}$}{$2^{++}$}}  &{Whole}              &{$F'$ wave}                      &{$P'$ wave}                      &{$D'$ wave}          \\ \hline
{Whole}                         &{73.9}               &{59.5}                           &{$3.4\times10^{-3}$}             &{0.620}              \\ \hline
{~~~~~$D$ wave~~~~~}            &{60.1}               &{57.1}                           &{$3.4\times10^{-4}$}             &{0.031}              \\ \hline
{$S$ wave}        &~~~~~{$1.9\times10^{-3}$}~~~~~ &~~~~~{$1.1\times10^{-5}$}~~~~~  &~~~~~{$5.3\times10^{-4}$}~~~~~   &~~~~~{$3.7\times10^{-5}$}~~~~~ \\ \hline
{$P$ wave}                      &{0.644}              &{0.029}                          &{$9.0\times10^{-4}$}             &{0.367}              \\ \hline
\end{tabular}
\end{center}
\end{table}

\subsection{Discussions about the charmonia}

\subsubsection{$\psi(2S)$ and $\psi(3S)$}
Their wave functions have similar partial wave content, a dominant $S$ partial wave and a small amount of $P$ and $D$ partial waves.
So they are the $S-P-D$ mixing states. In nonrelativistic limit, only the $S$ wave from $A_5$ and $A_6$ terms has a contribution; $P$ and $D$ waves and the $S$ wave from $A_3$ and $A_4$ terms provide the relativistic corrections.

\subsubsection{$\psi(3770)$ and $\psi(4160)$}
They are also the $S-P-D$ mixing states in our method, but in their wave functions, $D$ waves are dominant, mixed with a small amount of $S$ and $P$ waves. In the nonrelativistic limit, we have two choices: one is that only $D$ wave has contribution, and the other is that both the $S$ and $D$ waves have contributions. In the latter case, the $S$ wave from $A_3$ and $A_4$ terms and the $S$ wave from $A_5$ and $A_6$ terms both contribute significantly, but their contributions cancel out.

\subsubsection{$ \chi_{cJ}(1P)$ and $\chi_{cJ}(2P)$ ($J=0,1,2$)}
In our method, $\chi_{c0}(1P)$ and $\chi_{c0}(2P)$ are $P'$ partial wave dominant states with a small amount of $S'$ wave; they are $P'-S'$ mixing states. The $P'$ partial wave provides the nonrelativistic contribution, while the $S'$ wave gives the relativistic correction.
$\chi_{c1}(1P)$ and $\chi_{c1}(2P)$ are $P'-D'$ mixing states, where $D'$ wave gives the relativistic correction.

$\chi_{c2}(1P)$ and $\chi_{c2}(2P)$ are $P'$ wave dominant $P'-D'-F'$ mixing states. In the nonrelativistic limit, only the $P'$ wave from $F_5$ and $F_6$ terms has a contribution; $D'$ and $F'$ waves and the $P'$ wave from $F_3$ and $F_4$ terms provide the relativistic corrections.

\subsubsection{$\chi_{c2}(1F)$}

$\chi_{c2}(1F)$ is a $F'$ dominant $P'-D'-F'$ mixing state. In the nonrelativistic limit, similar to the cases of $\psi(3770)$ and $\psi(4160)$, we have two choices: one is that only $F'$ wave contributes, and the other is that both the $P'$ and $F'$ waves have contributions since the $P'$ wave from $F_3$ and $F_4$ terms and the $P'$ wave from $F_5$ and $F_6$ terms both contribute significantly, but their contributions also cancel out.

\subsection{Bottomonium's radiative decays}
At present, experiments have detected some bottomonia, and their masses are \cite{PDG}
\begin{eqnarray}
M_{{\Upsilon(2S)}}=&&10023~\rm{MeV}, ~M_{{\Upsilon(3S)}}=10355~\rm{MeV}, ~M_{{\chi_{_{b0}}(1P)}}=9859 ~\rm{MeV},
\nonumber\\
M_{{\chi_{_{b1}}(1P)}}=&&9893 ~\rm{MeV}, ~M_{{\chi_{_{b2}}(1P)}}=9912 ~\rm{MeV}, ~M_{{\chi_{_{b0}}(2P)}}=10233 ~\rm{MeV},
\nonumber\\
M_{{\chi_{_{b1}}(2P)}}=&&10255 ~\rm{MeV}, ~M_{{\chi_{_{b2}}(2P)}}=10269 ~\rm{MeV}.
\end{eqnarray}

In our model, the corresponding predicted masses are
\begin{eqnarray}
M_{{\Upsilon(2S)}}=&&10023~\rm{MeV}, ~M_{{\Upsilon(3S)}}=10369~\rm{MeV}, ~M_{{\chi_{_{b0}}(1P)}}=9859 ~\rm{MeV},
\nonumber\\
M_{{\chi_{_{b1}}(1P)}}=&&9892 ~\rm{MeV}, ~M_{{\chi_{_{b2}}(1P)}}=9912 ~\rm{MeV}, ~M_{{\chi_{_{b0}}(2P)}}=10241 ~\rm{MeV},
\nonumber\\
M_{{\chi_{_{b1}}(2P)}}=&&10273 ~\rm{MeV}, ~M_{{\chi_{_{b2}}(2P)}}=10289 ~\rm{MeV}.
\end{eqnarray}
It can be seen that the theoretical values agree much better with the experimental data than in the case of charmonium, but similar to the case of charmonium, we choose the experimental masses for calculation.

The $\Upsilon(1D)$, $\Upsilon(2D)$, and $\Upsilon(1F)$ have not been detected by the experiment. So in our calculation, their masses are taken from our previous study \cite{ret14},
$M_{_{\Upsilon(1D)}}=10130$ MeV, $M_{_{\Upsilon(2D)}}=10435$ MeV, and
$M_{_{\Upsilon(1F)}}=10372$ MeV. Our results of bottomonium radiative decays are shown in Table \ref{bottom}, for comparison. Theoretical results from other models and data from PDG are also shown in the same table.

Since the mass of the bottomonium is very heavy, the relativistic correction is small. Then from Table \ref{bottom}, we can see that, except for the channels $\Upsilon(3S)\to\chi_{_{bJ}}(1P)\gamma$, which have large uncertainties in theory, the results by most of the theoretical models are in agreement, at least comparable with each other, and also consist with data from PDG. For example, our results of $\Upsilon(2S)\to\chi_{_{bJ}}(1P)\gamma$ and $\Upsilon(3S)\to\chi_{_{bJ}}(2P)\gamma$ consist very well with experimental data and the theoretical predictions in Refs. \cite{ret17,ret38,ret39}.

Similar to the case of charmonium, because the contributions of wave functions on the two sides of the nodes strongly cancel each other out, the theoretical errors of processes $\Upsilon(3S)\to\chi_{_{bJ}}(1P)\gamma$ and $\Upsilon(2D)\to\chi_{_{bJ}}(1P)\gamma$, especially the former, are large, resulting in large differences between the results of different models.

\begin{table}[h]
\begin{center}
\caption{The decay widths (keV) of $\Upsilon\to\chi_{_{bJ}}\gamma$.}\label{bottom}
\begin{tabular}{c c c c c c c c c c}
\hline
\textbf{Process} &\textbf{Ours}&\textbf{\cite{ret41}}&\textbf{\cite{ret17}}&\textbf{\cite{ret42}}&\textbf{\cite{ret33}} &\textbf{\cite{ret38}}&\textbf{\cite{ret39}}&\textbf{\cite{ret35}}&\textbf{PDG}\cite{PDG}            \\
\hline
$\Upsilon(2S)\rightarrow\gamma\chi_{_{b0}}(1P)$     &1.13 &1.62 &1.29 &0.74 &1.19 &0.91&1.09 &1.09   &$1.22\pm0.23$   \\
$\Upsilon(2S)\rightarrow\gamma\chi_{_{b1}}(1P)$     &1.80 &2.45 &2.00 &1.40 &2.28 &1.63&1.84 &2.17   &$2.21\pm0.23$   \\
$\Upsilon(2S)\rightarrow\gamma\chi_{_{b2}}(1P)$     &1.83 &2.46 &2.04 &1.67 &2.58 &1.88&2.08 &2.62   &$2.29\pm0.30$   \\
\hline
$\Upsilon(1D)\rightarrow\gamma\chi_{_{b0}}(1P)$     &15.5 &23.4 &20.1 &12.5 &     &16.5&20.98&19.8   &                                 \\
$\Upsilon(1D)\rightarrow\gamma\chi_{_{b1}}(1P)$     &7.94 &12.7 &10.7 &7.59 &     &9.7 &12.29&13.3   &                                 \\
$\Upsilon(1D)\rightarrow\gamma\chi_{_{b2}}(1P)$     &0.416&0.69 &0.564&0.44 &     &0.56&0.65 &1.02   &                                 \\
\hline
$\Upsilon(3S)\rightarrow\gamma\chi_{_{b0}}(1P)$     &0.009&0.027&0.001&0.03 &0.12 &0.01&0.15 &0.097  &$0.055\pm0.012$ \\
$\Upsilon(3S)\rightarrow\gamma\chi_{_{b1}}(1P)$     &0.071&0.067&0.008&0.003&0.0  &0.05&0.16 &0.0005 &$0.018\pm0.012$ \\
$\Upsilon(3S)\rightarrow\gamma\chi_{_{b2}}(1P)$     &0.075&0.097&0.015&0.11 &0.20 &0.45&0.0827&0.14  &$0.203\pm0.039$ \\
$\Upsilon(3S)\rightarrow\gamma\chi_{_{b0}}(2P)$     &1.19 &1.49 &1.35 &1.07 &1.31 &1.03&1.21  &3.330 &$1.20\pm0.23$   \\
$\Upsilon(3S)\rightarrow\gamma\chi_{_{b1}}(2P)$     &2.18 &2.41 &2.20 &2.05 &2.66 &1.91&2.13  &2.61  &$2.56\pm0.48$   \\
$\Upsilon(3S)\rightarrow\gamma\chi_{_{b2}}(2P)$     &2.50 &2.67 &2.40 &2.51 &3.18 &2.30&2.56  &3.16  &$2.66\pm0.57$   \\
\hline
$\Upsilon(2D)\rightarrow\gamma\chi_{_{b0}}(1P)$     &2.16 &     &3.60 &     &     &2.9 &3.52  &5.56  &                                 \\
$\Upsilon(2D)\rightarrow\gamma\chi_{_{b1}}(1P)$     &1.70 &     &     &     &     &0.9 &1.58  &2.17  &                                 \\
$\Upsilon(2D)\rightarrow\gamma\chi_{_{b2}}(1P)$     &0.091&     &     &     &     &0.02&0.0608&0.44  &                                 \\
$\Upsilon(2D)\rightarrow\gamma\chi_{_{b0}}(2P)$     &12.5 &     &13.1 &     &     &10.6&8.35  &9.58  &                                 \\
$\Upsilon(2D)\rightarrow\gamma\chi_{_{b1}}(2P)$     &6.14 &     &     &     &     &6.5 &4.84  &6.74  &                                 \\
$\Upsilon(2D)\rightarrow\gamma\chi_{_{b2}}(2P)$     &0.371&     &     &     &     &0.4 &0.24  &0.47  &                                 \\
$\Upsilon(2D)\rightarrow\gamma\chi_{_{b2}}(1F)$     &0.379&     &0.833&     &     &1.6 &2.05  & &                                      \\
\hline
\end{tabular}
\end{center}
\end{table}

\subsection{Contributions of different partial waves in bottomonia decays}
For the decays of bottomonia, which have the same quantum numbers with charmonia, the
calculations are similar, and we will not repeat them one by one, but focus on some different
contents. The bottomonium is much heavier than charmonium, so the relativistic correction of bottomonium is much smaller than that of charmonium. This leads to the fact that the content of the small partial wave in bottomonium is much smaller than the corresponding charmonium case; see Tables \ref{spd},\ref{ps},\ref{pdf} for details.
\subsubsection{$\Upsilon(2S) \to \chi_{b0}(1P)\gamma$}
As expected, we note that in Table \ref{bottom0+}, the small component terms, namely, the $P$ wave and $D$ wave of $\Upsilon(2S)$, and $S'$ wave of $\chi_{b0}(1P)$ make very little contribution to the decay $\Upsilon(2S) \to \chi_{b0}(1P)\gamma$. The relativistic effect,
$$\frac{\Gamma_{rel}-\Gamma_{non-rel}}{\Gamma_{rel}}=4.43\%,$$
is very small.

\begin{table}[h]
\begin{center}
\caption{Contributions of different partial waves to the decay width (keV) of $\Upsilon(2S) \to \chi_{_{b0}}(1P)\gamma$.}\label{bottom0+}
{\begin{tabular}{|c|c|c|c|}
\hline
\diagbox{$1^{--}$}{$0^{++}$}   &~~ Whole  ~~                   & $P'$ wave                            & $S'$ wave             \\ \hline
  Whole                        & 1.13                          & 1.08                                 & $6.9\times 10^{-4}$   \\ \hline
  $S$ wave                     & 1.08                          & 1.08                                 & 0                     \\ \hline
  $P$ wave                     & $5.8\times 10^{-4}$           & $5.2\times 10^{-6}$                  & $7.0\times 10^{-4}$   \\ \hline
  ~~~~~~~$D$ wave~~~~~~~  & ~~~~~$1.8\times 10^{-7}$~~~~~ & ~~~~~$1.2\times 10^{-7}$~~~~~ & ~~~~~$5.7\times 10^{-9}$~~~~~     \\ \hline
\end{tabular}}
\end{center}
\end{table}

\subsubsection{$\Upsilon(2S) \to \chi_{b1}(1P)\gamma$}
Some details of the channel $\Upsilon(2S) \to \chi_{b1}(1P)\gamma$ are listed in Table \ref{bottom1+}. We can see that, similar to the process $\Upsilon(2S) \to \chi_{b0}(1P)\gamma$, the nonrelativistic result plays a major role in this process. The relativistic effect of this process is
$$\frac{\Gamma_{rel}-\Gamma_{non-rel}}{\Gamma_{rel}}=6.90\%.$$

\begin{table}[h]
\begin{center}
\caption{Contributions of different partial waves to the decay width (keV) of $\Upsilon(2S) \to \chi_{_{b1}}(1P)\gamma$.}\label{bottom1+}
{\begin{tabular}{|c|c|c|c|}
\hline
\diagbox{$1^{--}$}{$1^{++}$} & ~~Whole   ~~                   & $P'$ wave ($D_1$,$D_2$)       & $D'$ wave ($D_3$)         \\ \hline
  Whole                      & 1.80                           & 1.67                          & $2.2\times 10^{-3}$       \\ \hline
  $S$ wave                   & 1.73                           & 1.67                          & $5.7\times 10^{-4}$       \\ \hline
  $P$ wave                   & $4.9\times 10^{-4}$            & $1.2\times 10^{-6}$           & $5.3\times 10^{-4}$       \\ \hline
  ~~~~~~~$D$ wave~~~~~~~ & ~~~~~$1.3\times 10^{-6}$~~~~~ & ~~~~~$2.0\times 10^{-6}$~~~~~ & ~~~~~$8.6\times 10^{-8}$~~~~~  \\ \hline
\end{tabular}}
\end{center}
\end{table}

\subsubsection{$\Upsilon(2S) \to \chi_{b2}(1P)\gamma$}
\begin{table}[htb]
\begin{center}
\caption{Contributions of different partial waves to the decay width (keV) of $\Upsilon(2S) \to \chi_{_{b2}}(1P)\gamma$.}\label{bottom2+}
{\begin{tabular}{|c|c|c|c|c|}
\hline
\diagbox{$1^{--}$}{$2^{++}$}    & Whole                        & $P'$ wave                    & $D'$ wave                    & $F'$ wave            \\ \hline
  Whole                         &1.83                          &1.79                          &$2.4\times 10^{-4}$           &$1.2\times 10^{-7}$   \\ \hline
  $S$ wave                      &1.77                          &1.74                          &$1.2\times 10^{-4}$           &$5.5\times 10^{-9}$   \\ \hline
  $P$ wave                      &$5.6\times 10^{-4}$           &$3.5\times 10^{-4}$           &$1.7\times 10^{-5}$           &$1.4\times 10^{-8}$   \\ \hline
  ~~~~~$D$ wave~~~~~       &~~~~~$5.1\times 10^{-8}$~~~~~ &~~~~~$1.1\times 10^{-7}$~~~~~ &~~~~~$8.0\times 10^{-7}$~~~~~ &~~~~~$1.8\times 10^{-7}$~~~~~            \\ \hline
\end{tabular}}
\end{center}
\end{table}

In the process $\Upsilon(2S) \to \chi_{b2}(1P)\gamma$, see Table \ref{bottom2+} for details, the main contribution to the decay width comes from the main partial waves of the initial and final states, that is, from the interaction of $S\times P'$. And the relativistic effect in this process is $4.89\%$.
\subsubsection{$\Upsilon(1D) \to \chi_{b0}(1P)\gamma$}
Similar to $\psi(3770)$, $\Upsilon(1D)$ is a $D$ wave dominant $S-P-D$ mixing state.
Table \ref{bb0+} shows some details of the decay $\Upsilon(1D) \to \chi_{b0}(1P)\gamma$; we can see that the contribution of small partial waves of bottomonia is much smaller than that of the corresponding charmonia. If the relativistic corrections comes from the $P$ wave in $\Upsilon(1D)$ and the $S'$ wave in $\chi_{b0}(1P)$, the relativistic effect is
$$\frac{\Gamma_{rel}-\Gamma_{non-rel}}{\Gamma_{rel}}=3.70\%.$$
If both $S$ and $P$ waves in $\Upsilon(1D)$ are relativistic corrections, then the relativistic effect is also $3.75\%$.

\begin{table}[h]
\begin{center}
\caption{Contributions of different partial waves to the decay width (keV) of $\Upsilon(1D) \to \chi_{_{b0}}(1P)\gamma$.}\label{bb0+}
{\begin{tabular}{|c|c|c|c|}
\hline
\diagbox{$1^{--}$}{$0^{++}$}  & Whole                         & $P'$ wave ($B_1$,$B_2$)             & $S'$ wave ($B_3$)             \\ \hline
  Whole                       & 15.5                          & 15.0                                & $4.1\times 10^{-3}$           \\ \hline
  ~~~~~~~$D$ wave~~~~~~~      & 15.0                          & 14.9                           & ~~~~~$1.2\times 10^{-6}$~~~~~      \\ \hline
  $S$ wave                    & ~~~~~$1.2\times 10^{-6}$~~~~~ & ~~~~~$1.2\times 10^{-4}$~~~~~       & 0                             \\ \hline
  $P$ wave                    &~~ $3.9\times 10^{-3}$ ~~      & $8.3\times 10^{-5}$                 & $2.9\times 10^{-3}$           \\ \hline
\end{tabular}}
\end{center}
\end{table}
\subsubsection{$\Upsilon(1D) \to \chi_{b1}(1P)\gamma$}
From Table \ref{bb1+}, it can be seen that, although there are some differences, we can get the same conclusion of $\Upsilon(1D) \to \chi_{_{b1}}(1P)\gamma$ as in the process $\Upsilon(1D) \to \chi_{b0}(1P)\gamma$. The relativistic effects are $7.33\%$ and $7.21\%$ in the two choices, similar with $\Upsilon(1D) \to \chi_{b0}(1P)\gamma$.

\begin{table}[h]
\begin{center}
\caption{Contributions of different partial waves to the decay width (keV) of $\Upsilon(1D) \to \chi_{_{b1}}(1P)\gamma$.}\label{bb1+}
{\begin{tabular}{|c|c|c|c|}
\hline
\diagbox{$1^{--}$}{$1^{++}$} &~~ Whole ~~                   & $P'$ wave ($D_1$,$D_2$)  & $D'$ wave ($D_3$)         \\ \hline
  Whole                      & 7.94                         & 7.99                     & $8.5\times 10^{-5}$       \\ \hline
  ~~~~~~~$D$ wave~~~~~~~     &7.31                          &7.37                      & $1.4\times 10^{-4}$       \\ \hline
  $S$ wave       &~~~~~$2.4\times 10^{-7}$~~~~~ &~~~~~$3.0\times 10^{-6}$~~~~~     &~~~~~$4.9\times 10^{-6}$~~~~~  \\ \hline
  $P$ wave                   & 0.0132                       & 0.0131                   & $2.9\times 10^{-7}$       \\ \hline
\end{tabular}}
\end{center}
\end{table}
\subsubsection{$\Upsilon(1D) \to \chi_{b2}(1P)\gamma$}
This process is similar to the decay $\psi(3770) \to \chi_{c2}(1P)\gamma$, but with a small relativistic effect; see Table \ref{bb2+} for details. With the same two choices with $\psi(3770) \to \chi_{c2}(1P)\gamma$, the relativistic effects are $3.19\%$ and $3.86\%$.

\begin{table}[h]
\begin{center}
\caption{Contributions of different partial waves to the decay width (keV) of $\Upsilon(1D) \to \chi_{_{b2}}(1P)\gamma$.}\label{bb2+}
{\begin{tabular}{|c|c|c|c|c|}
\hline
\diagbox{$1^{--}$}{$2^{++}$}   & Whole                   & $P'$ wave ($F_5$,$F_6$)        & $D'$ wave ($F_1$,$F_2$,$F_7$)& $F'$ wave ($F_3$,$F_4$)   \\ \hline
  Whole                        & {0.416}                        &{0.443}                  & {$6.6\times 10^{-4}$}       & {$2.0\times 10^{-6}$}      \\ \hline
  ~~~~~$D$ wave~~~~~           &{0.387}                         &{0.396}                  & {$9.3\times 10^{-5}$}       & {$1.5\times 10^{-6}$}      \\ \hline
  $S$ wave         &~~~~~{$2.2\times 10^{-6}$}~~~~~     &{$4.2\times 10^{-6}$} & ~~~~~{$1.2\times 10^{-8}$}~~~~~        & {$3.4\times 10^{-10}$}     \\ \hline
  $P$ wave            &~~{$5.2\times 10^{-4}$}~~   &~~~~~{$1.2\times 10^{-3}$}~~~~~       & {$2.6\times 10^{-4}$}  & ~~~~~{$4.4\times 10^{-8}$}~~~~~ \\ \hline
\end{tabular}}
\end{center}
\end{table}
\subsubsection{$\Upsilon(2D) \to \chi_{b2}(1F)\gamma$}
$\Upsilon(2D)$ is a $S-P-D$ mixing state dominated by $2D$ wave. Its radiative decay to the final state $\chi_{bJ}(1P)$ or $\chi_{bJ}(2P)$ ($J=0,1,2$) has many similarities with $\Upsilon(1D) \to \chi_{bJ}(1P)\gamma$, and they belong to the same type of process. So we will not give details about them, only show the details of the decay $\Upsilon(2D) \to \chi_{b2}(1F)\gamma$ in Table \ref{bb2d}, where $\chi_{b2}(1F)$ is the $1F'$ dominant $P'-D'-F'$ mixing state. $\chi_{b2}(1F)$ has not been detected by experiment, and we predict its mass to be about $10374$ MeV \cite{ret14}.

The radiative decay $\Upsilon(2D) \to \chi_{b2}(1F)\gamma$ is similar to $\psi(4160)\to \chi_{c2}(1F)\gamma$; the difference is that the relativistic correction of bottomonium is very small due to its heavy mass, see Table \ref{bb2d} for details. If we choose the same conditions, the relativistic effects are calculated as $4.68\%$ and $4.66\%$.

\begin{table}[h]
\begin{center}
\caption{Contributions of different partial waves to the decay width (keV) of $\Upsilon(2D) \to \chi_{_{b2}}(1F)\gamma$. }\label{bb2d}
\begin{tabular}{|c|c|c|c|c|}
\hline
{\diagbox{$1^{--}$}{$2^{++}$}} &{Whole}                   &{$F'$ wave}                &{$P'$ wave}                      &{$D'$ wave}                   \\ \hline
{Whole}                        &{0.379}                   &{0.366}                    &{$2.2\times 10^{-4}$}            &{$1.3\times10^{-4}$}          \\ \hline
{~~~~~$D$ wave~~~~~ }          &{0.371}                   &{0.361}                    &{$2.2\times 10^{-4}$}            &{$6.2\times 10^{-5}$}         \\ \hline
{$S$ wave }&~~~~~{$2.1\times 10^{-7}$}~~~~~     &~~~~~{$1.5\times 10^{-7}$}~~~~~ &~~~~~{$5.5\times 10^{-8}$}~~~~~  &~~~~~{$3.4\times 10^{-9}$}~~~~~    \\ \hline
{$P$ wave}                     &{$6.0\times10^{-5}$} &{$1.8\times 10^{-5}$}           &{$5.8\times 10^{-8}$}            &{$2.5\times10^{-5}$}          \\ \hline
\end{tabular}
\end{center}
\end{table}

\subsection{Discussions about the bottomonia}
\subsubsection{$\Upsilon(2S)$ and $\Upsilon(3S)$}
These two states are similar to $\psi(2S)$ and $\psi(3S)$; their wave functions are both $S$ wave dominant with small amount of $P$ and $D$ waves, so they are $S-P-D$ mixing states. Compared with charmonium, The relativistic correction of bottomonium is much smaller, which indicates that the nonrelativistic approximation is good for bottomonium.

\subsubsection{$\Upsilon(1D)$ and $\Upsilon(2D)$}
These two states are also $S-P-D$ mixing states. In their wave functions, $D$ wave is the dominant one, $S$ and $P$ waves have a small proportion. However, the $S$ wave has two sources, both of which have a large proportion when they exist separately, while combined, the total proportion of $S$ wave is small.

\subsubsection{$ \chi_{bJ}(1P)$ and $\chi_{bJ}(2P)$ ($J=0,1,2$)}
$\chi_{bJ}(1P)$ and $\chi_{bJ}(2P)$ ($J=0,1,2$) are all $P'$ wave dominant states. Among them, $\chi_{b0}(1P)$ and $\chi_{b0}(2P)$ are $P-S$ mixing states, where the $S$ wave provides the relativistic correction. $\chi_{b1}(1P)$ and $\chi_{b1}(2P)$ are $P-D$ mixing states, and the $D$ wave provides the relativistic correction. While $\chi_{b2}(1P)$ and $\chi_{b2}(2P)$ are $P-D-F$ mixing states, where the $D$ wave alone can be considered as the relativistic correction, or the $D$ wave and $F$ wave, as well as the $P$ wave from $F_3$ and $F_4$ terms, can be considered as the relativistic correction.

\subsubsection{$\chi_{b2}(1F)$}
$\chi_{b2}(1F)$ is also a typical $P-D-F$ mixing state, but its wave function is dominated by the $F$ wave and contains small component of $P$ and $D$ waves. Where the $P$ wave has two sources, both of which have a large proportion separately, while their sum is small.

\section{Summary}
We confirm that the wave functions of mesons are not composed of a single pure wave and contain other different partial waves, so all the mesons are mixing states.
We study the partial waves of heavy quarkonium and their contributions to radiative electromagnetic decay. The results show that for the $S$ and $P$ wave dominated states, for example, $\psi(nS)$, $\Upsilon(nS)$ ($n=2,3$), $\chi_{_{cJ}}(mP)$, and $\chi_{_{bJ}}(mP)$ ($m=1,2;J=0,1,2$), the dominant $S$ and $P$ waves provide the main and nonrelativistic contribution, while the partial waves of the small components mainly contribute to the relativistic correction.

$\psi(nD)$ and $\Upsilon(nD)$ ($n=1,2$) are $D$ wave dominant $S-P-D$ mixing states, and $\chi_{c2}(1F)$ and $\chi_{b2}(1F)$ are $F$ wave dominant $P-D-F$ mixing states. But their wave functions are complicated; we cannot simply only treat the dominant $D$ wave or $F$ wave as the nonrelativistic contribution, since the $S$ wave in $S-P-D$ mixing state or $P$ wave in $P-D-F$ mixing state has two sources, and both sources have large proportions separately, while their contributions cancel out and are not significant in this article.

Our results of charmonium electromagnetic decay are comparable with the experimental data, and the results of bottomonium are in good agreement with the existing data. We calculate the radiative decays of the mixed states and find that the $\psi(2D)\to \chi_{cJ}(2P)$, $\Upsilon(1D)\to \chi_{bJ}(1P)$, and $\Upsilon(2D)\to \chi_{bJ}(2P)$ ($J=0,1$) transitions have large partial decay widths, may be helpful to find these undiscovered particles.

{\bf Acknowledgments}
This work was supported in part by the National Natural Science Foundation of China (NSFC) under the Grants No. 12075073 and No. 11865001, the Natural Science Foundation of Hebei province under the Grant No. A2021201009, and Post-graduate's Innovation Fund Project of Hebei University under the Grant No. HBU2022BS002.


 \end{document}